\documentclass[aps,pre,reprint,showpages,amsmath,preprintnumbers,nofootinbib]{revtex4-1}
\usepackage[dvipdfmx,hiresbb]{graphicx} 
\usepackage{graphicx}
\usepackage{comment}
\usepackage{url}
\usepackage{amsmath}
\usepackage{color}
\usepackage{natbib}

\begin{document}
\preprint{
KEK-TH 1715
}
\date{\today}
\title{Heat capacity of liquids in light of hydrodynamics as U(1) gauge theory}

\author{Yoshinori Tomiyoshi}
\affiliation {Department of Physics, Graduate School of Science, Tohoku University, Sendai 980-8578, Japan}

\author{Daiki Ueda}
\affiliation{
KEK Theory Center, IPNS, Ibaraki 305-0801, Japan
}
\affiliation{
The Graduate University of Advanced Studies (Sokendai), Tsukuba, Ibaraki 305-0801, Japan}

\begin{abstract}
We investigate the heat capacity of liquids through a theoretical approach based on a quasiparticle description. 
By interpreting the microscopic dynamics of particles in liquids in terms of quasiparticles, 
we suggest a simplified understanding of the number of degrees of freedom in liquids.
A equivalence between hydrodynamics and U(1) gauge theory, which is newly proposed in the present paper,  develops the quasiparticle description to construct a new Lagrangian which correctly reproduces the number of modes at the melting points and at the critical points. The heat capacity evaluated from this Lagrangian naturally interpolates between these two points, and agrees with the phonon theory of liquids [Sci. Rep. \textbf{2}, 421 (2012)].

\end{abstract}
\pacs{
05.70.Ln
}

\maketitle

\section{Introduction}
The liquid phase of matter still remains one of the most challenging subjects in statistical mechanics due to its irregular structure contrary to the solid phase and its strong intermolecular interactions unlike the gas phase \cite{Egelstaff_1992_textbook,Hansen_1986_textbook}.
As opposed to the solid state where the statistical mechanics of the Debye model correctly predicts thermodynamical properties, the intermediate nature of the liquid state prevents us from constructing a general model for liquids.
The heat capacity of a liquid at constant volume, which decreases in general from about $3Nk_{\mathrm{B}}$ at the melting point to about $2Nk_{\mathrm{B}}$ at the critical temperature \cite{Faber_1972_textbook,Grimvall_1975,Grimvall_1996,Grimvall_2005}. It has not been understood adequately compared with solid, although it is an important thermodynamical property to specify the degrees of freedom of the system. 
In spite of these difficulties, a phonon theory of liquid was proposed \cite{Trachenko_2008, Bolmatov_2011,Bolmatov_2012,Trachenko_2016}, which quantitatively reproduces the temperature dependence of heat capacities in various liquids \cite{Trachenko_2008,Bolmatov_2011,Bolmatov_2012, Trachenko_2016,YongKang_2014}.
This theory is based on a physical insight into the microscopic dynamics of particles in liquids referred to as Frenkel's idea \cite{Trachenko_2016} that a particle in liquids oscillates around a local stable region during a characteristic time $\tau_{\mathrm{F}}$, and after that the particle can escape from the local stable region \cite{Frenkel_1946_textbook}.    
This idea allows us to define a characteristic frequency $\omega_{\mathrm{F}}=1/\tau_{\mathrm{F}}$, which leads to a classification of the dynamics of liquids into two regimes:
one is the solid-like regime $(\omega > \omega_{\mathrm{F}})$ where liquids support one longitudinal and two transverse phonon modes, and the other is the hydrodynamic regime $(\omega < \omega_{\mathrm{F}})$ where liquids retain only the longitudinal mode \cite{Trachenko_2008, Bolmatov_2011,Bolmatov_2012, Trachenko_2016, Frenkel_1946_textbook}.
Following this idea, the authors \cite{Bolmatov_2012, Trachenko_2008,Bolmatov_2011,Trachenko_2016} estimate the total energy of liquids, from which they give an analytical expression for the heat capacity. 
The authors, however, neglect the contribution from the diffusive jump to the total energy, although such a process is regarded as one of the main dynamical modes in their qualitative explanation \cite{Trachenko_2008,Bolmatov_2011,Bolmatov_2012,Trachenko_2016}. 
The dynamics of particles in liquids is described by a combination of the oscillation and the diffusive jump. 
These dynamical processes cannot be microscopically tractable since we face a complicated many-body problem due to their intermolecular interactions.
From the perspective of modern condensed matter physics, quasiparticles or elementary excitations play an important role in the description of strongly correlated systems \cite{Altland_2010_textbook}.
For liquids, the existence of an elementary excitation different from phonon was recently suggested \cite{Egami_2007,Iwashita_2013}, which motivates a theoretical approach to the viscosity of liquids \cite{Bellissard_2017,Bellissard_2018}.
It is well known that the heat capacity of solids reflects the number of degrees of freedom associated with the quasiparticles, e.g. phonons.    
Then, the quasiparticle description of liquids is expected to be a key to construct a general theory for heat capacity of liquids. 
To realize this strategy in a natural manner, we turn our attention to the correspondence between hydrodynamics and electromagnetism whose clarifications and employments have been done by several researchers \cite{Marmanis_1998, Mendes_2003, Kambe_2010, Abreu_2015}.
This correspondence suggests an application of statistical mechanics on electromagnetism to the study of the heat capacity of liquids. 
Stimulated by the previous studies, we propose a theoretical approach to the heat capacity of liquids from a standpoint of continuum mechanics.
We show that there exist quasiparticles inherent in the hydrodynamic Euler equation, and that these quasiparticles correspond to the diffusive motion of particles and to the density fluctuations in liquids.
These quasiparticles are naturally incorporated into a hydrodynamic Lagrangian through an equivalence between hydrodynamics and U(1) gauge theory.
Such an equivalence is newly proposed in the present paper. 
Moreover, to take into account the solid-like behavior in the high frequency regime  \cite{Trachenko_2016, Frenkel_1946_textbook,Trachenko_2008,Bolmatov_2012, Bolmatov_2011}, we also propose a new Lagrangian for liquids, which is composed of the hydrodynamic and the elastic contributions in the lower and higher frequency regimes than a characteristic frequency.
The transformation between hydrodynamics and elasticity can be performed by the arbitrariness in the choice of gauge-fixing conditions. 
The heat capacity calculated from our Lagrangian with an appropriate temperature dependence of the characteristic frequency $\omega_{\mathrm{F}}$ agrees with those of the phonon theory of liquid \cite{Trachenko_2008, Bolmatov_2011,Bolmatov_2012,Trachenko_2016}.

\section{Microscopic dynamics in liquids}
\begin{figure}[t]
\begin{center}
\includegraphics[width=.45\textwidth]{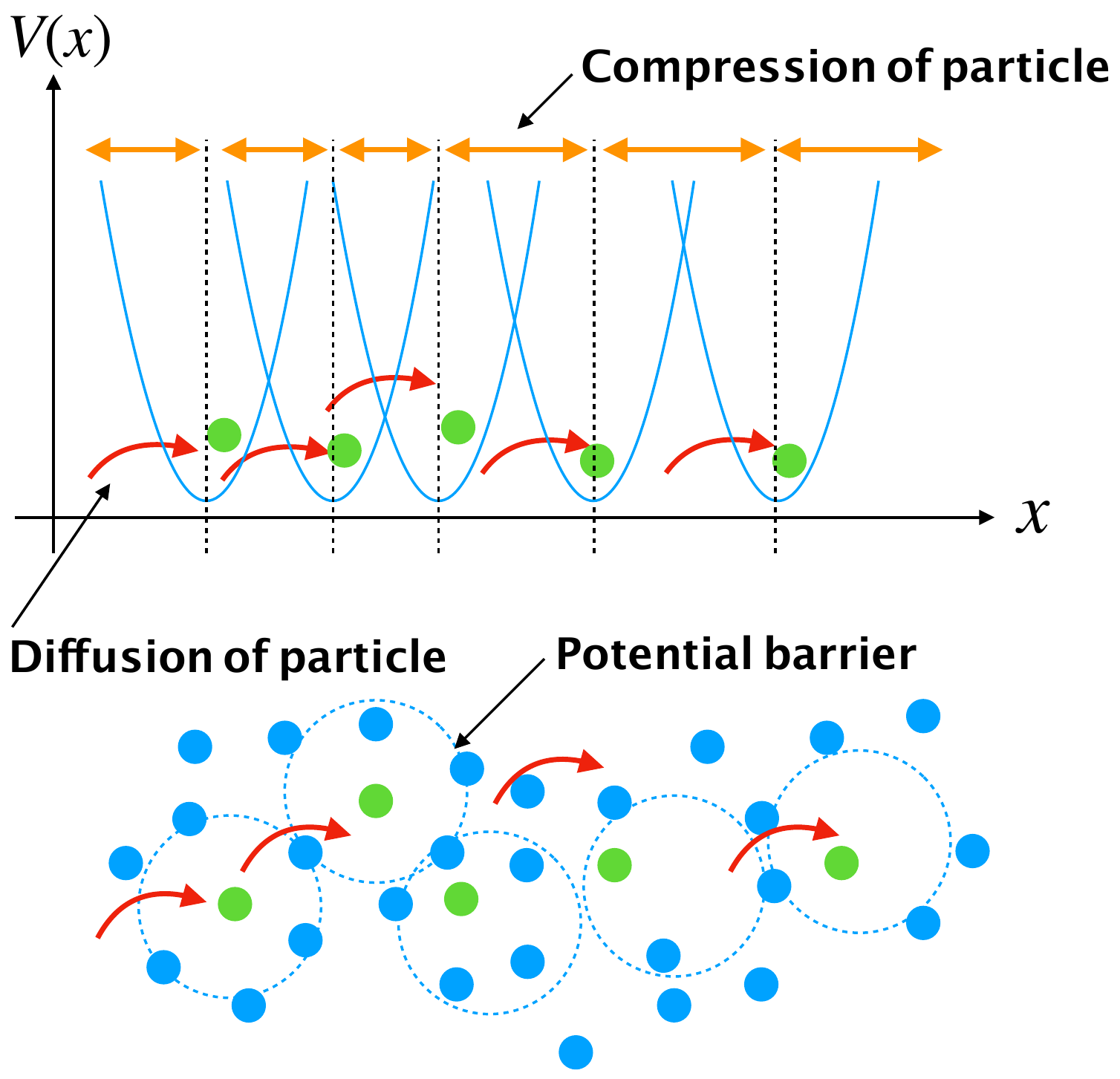}
\caption{A schematic pictures of the microscopic dynamics of particles in liquids.  
In the shorter time regime than $\tau_{\mathrm{F}}$, a tagged (green) particle oscillates in the potential barrier formed by the surrounding (blue) particles.
In the longer time regime than $\tau_{\mathrm{F}}$, on the other hand, the tagged particle moves to one of the neighboring local minima through the diffusive motion.  
In both of these time regimes, a density fluctuation of surrounding particles occurs. This mode is intuitively understood as a change in the distances of local minima of the potential. 
}
\label{fig:sc1}
\end{center}
\end{figure}

\begin{figure}[t]
\begin{center}
\includegraphics[width=.48\textwidth]{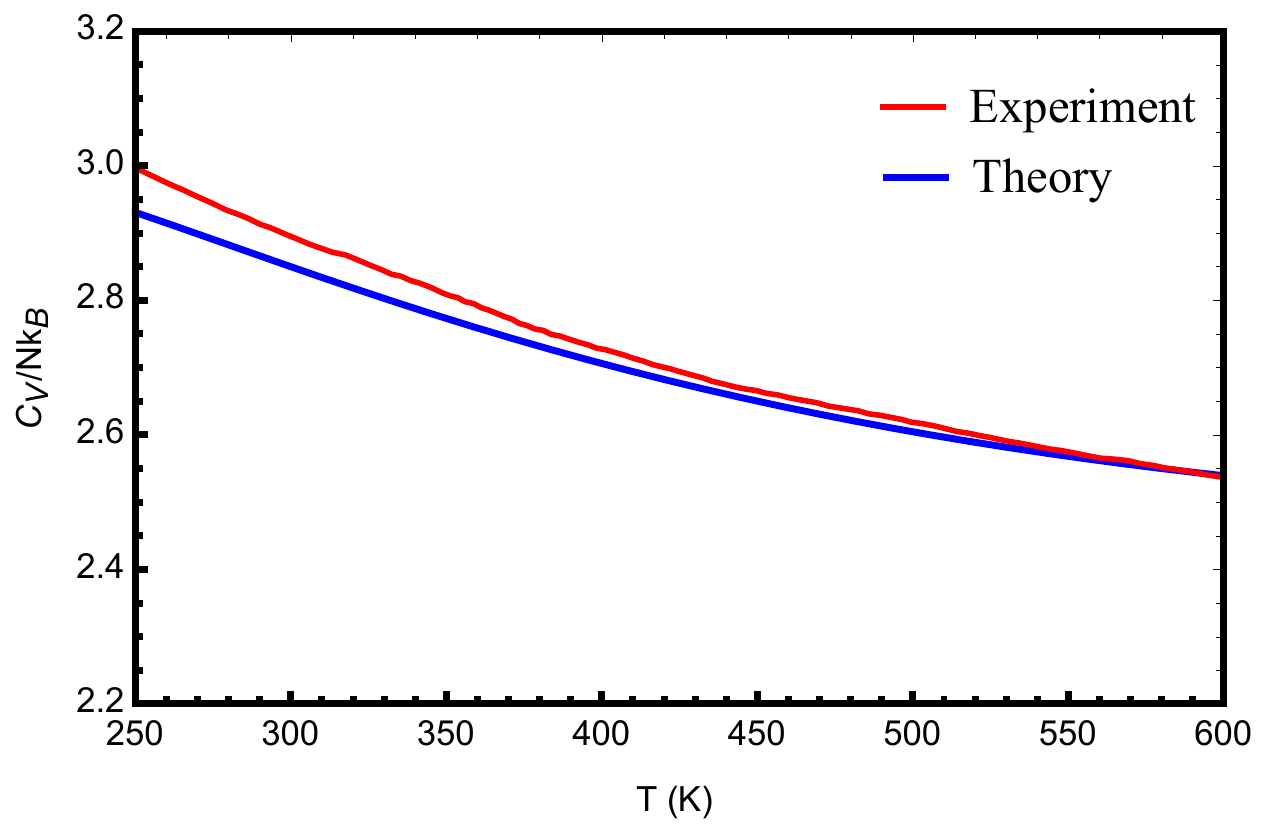}
\caption{Heat capacity of liquid mercury per atom as a function of temperature.
The red curve is the experimental data in Ref. \cite{Wallace_1998}. 
The blue curve is calculated by using Eq.(\ref{eq: enel_first}).
The set of parameters used here is based on \cite{Trachenko_2008}.
The deviation the experimental data from the theoretical prediction is within the experimental error shown in Ref. \cite{Wallace_1998}.
}
\label{fig:plot}
\end{center}
\end{figure}

In this section, we begin by reviewing in detail the qualitative picture of the microscopic dynamics of particles in liquids explained in Refs. \cite{Trachenko_2016, Frenkel_1946_textbook,Trachenko_2008, Bolmatov_2011}. 
Following this picture, we clarify the relationship between the dynamics of particles in liquids and the dynamical modes retained in the continuum mechanical equations, that is, the elastic and hydrodynamic equations.
In liquids, a particle experiences many-body interactions between neighboring particles due to the van der Waals force or the Coulomb force \cite{Egelstaff_1992_textbook} and the mean free path is the same order of the particle spacing in solids \cite{Takagi_1980}, which results in the complicated dynamical correlations in liquids.
According to Refs. \cite{Trachenko_2016, Frenkel_1946_textbook,Trachenko_2008, Bolmatov_2011,Bolmatov_2012}, under such an environment, a tagged particle undergoes the solid-like oscillatory motion in the effective potential barriers formed by the surrounding particles.
However, the tagged particle is not permanently trapped in the potential, and undertakes the hydrodynamic diffusive motion by which the tagged particle escapes from the potential well on average with a characteristic time scale $\tau_{\mathrm{F}}$.
In addition, the distances between local minima of the potential also fluctuate as a consequence of the density fluctuation of the surrounding particles.
These motions are illustrated in Fig.\ref{fig:sc1}.
The above picture is based on a single characteristic relaxation time for liquids, which leads to viscoelastic model of liquids.
We can identify $\tau_{\mathrm{F}}$ as the Maxwell relaxation time $\tau_{\mathrm{M}}=\eta/G_{\infty}$, where $\eta$ is the viscosity and $G_{\infty}$ is the high-frequency shear modulus.  
The dynamical property of liquids is explained by the combination of these modes, and the value of $\tau_{\mathrm{F}}$ decides the relative weight of each of these two contributions, which is a function of temperature.
%
%
In order to make our discussion clearer, we summarize the dynamics of particles in each frequency regime separated by the characteristic frequency $\omega_{\mathrm{F}}=1/\tau_{\mathrm{F}}$: 
\begin{enumerate}
\item In the higher frequency regime $\omega > \omega_{\rm F}$, which we refer to as the elastic regime, the tagged particle  oscillates in the potential formed by the surrounding particles. 
\item In the lower frequency $\omega < \omega_{\rm F}$, which we refer to as the hydrodynamic regime, the tagged particle escapes from the potential well via the hydrodynamic diffusive motion, and the surrounding particles show the density fluctuation as collective motions.
\end{enumerate}
Now we propose the quasiparticle description of these motions. 
The result presented in the following is basically same as the phonon theory of liquid \cite{Bolmatov_2012,Bolmatov_2011,Trachenko_2016}, but different from their theory especially in the interpretation of the hydrodynamic regime.
In the elastic regime, the motion of particles are correctly captured by the same mechanism of the elastic lattice vibration in the Debye model, which produces two transverse and one longitudinal modes of phonons.
Therefore, the dynamics of particles clearly obeys the elastic wave equations as
\begin{align}
\left(\frac{1}{c_{\mathrm{s}}^2} \frac{\partial^2}{\partial t^2} -\nabla^2 \right)\vec{u}_m=0, 
\end{align}
where $c_{\mathrm{s}}$ is the sound speed, $m={\rm L} ~ {\rm or} ~ {\rm T}$, L and T being the longitudinal and transverse modes. 
In the hydrodynamic regime, it is important to microscopically understand the dynamical modes retained in the hydrodynamic equations.
To simplify the problem, we adopt the ideal fluid. 
In hydrodynamics, the equation of motion of the ideal fluid is expressed by the Euler equation 
\begin{align}
	\frac{\partial}{\partial t}\vec{v} + (\vec{v}\cdot \nabla) \vec{v} = -\frac{1}{\rho}\nabla p,\label{eq:Euler} 
\end{align}
and auxiliary conditions such as the equation of continuity of density
\begin{align}
	\frac{\partial}{\partial t}\rho + \vec{v}\cdot \nabla \rho + \rho \nabla \cdot \vec{v} = 0,\label{eq:continuity_on_density}
\end{align}
where $\vec{v}$ is the fluid velocity, $\rho$ is the fluid density and $p$ is the pressure.
If we assume an isentropic flow, we can utilize a simple thermodynamical relation among the deviations of enthalpy $h$ per a mass, pressure $p$ and density $\rho$ from their equilibrium values  \cite{Landau_1969_statistical1,Kambe_2010} as 
\begin{align}
	\Delta h =\frac{1}{\rho}\Delta p= \frac{c_{\rm s}^2}{\rho}\Delta \rho,
	\label{eq:enthalpy_pressure_relation}
\end{align}
In the following, the variables stand for the deviations from their equilibrium values. 
By using Eq. (\ref{eq:enthalpy_pressure_relation}) and linearizing Eqs. (\ref{eq:Euler}) and (\ref{eq:continuity_on_density}) with respect to $\vec{v}$ and $h$, we finally obtain the linearized Euler equation as
\begin{align}
	\frac{\partial}{\partial t}h + c_{\rm s}^2 \nabla\cdot \vec{v} = 0, \label{eq:equation_of_h} \\
	\frac{\partial}{\partial t}\vec{v} + \nabla h = 0. \label{eq:equation_of_v}
\end{align}
In Eqs. (\ref{eq:equation_of_h}) and (\ref{eq:equation_of_v}), there exist two dynamical independent modes:
the one is  the enthalpy field $h$ and the other is the velocity potential $\Phi$, which is defined by $\vec{v}= \nabla \Phi$ \cite{Landau_fluid}.
The vorticity is now conserved through Eq. (\ref{eq:equation_of_v}).
The enthalpy field reflects the density fluctuation via Eq. (\ref{eq:enthalpy_pressure_relation}), which corresponds to the fluctuation of distances of local minima of the potential.
The velocity potential, on the other hand, represents the diffusive flow of particles in and out of the local region, which corresponds to the diffusion motion among the local minima of potential.
These quasiparticles can be associated with a gauge field in U(1) gauge theory in the next section. 
In the above discussion, we clarified the number of modes in each regime: three modes in the elastic regime and two modes in the hydrodynamic regime. 
This fact enables us to immediately guess a form of the partition function in each regime: 
\begin{enumerate}
	\item In the elastic regime, 
	\begin{align}
	Z_{\rm total}^{\rm e} = \left( \prod_{\vec{k}}e^{-\beta \hbar \omega_{\vec{k}} } \right)^3~{\rm for}~\vec{k}~s.t.~ \omega_{\rm F}<  \omega_{\vec{k}} < \omega_{\rm D}.
	\label{eq: part1}
\end{align}
	\item  In the hydrodynamic regime,
	 \begin{align}
	Z_{\rm total}^{\rm h} =\left( \prod_{\vec{k}}e^{-\beta  \hbar \omega_{\vec{k}} } \right)^2~{\rm for}~\vec{k}~s.t.~0< \omega_{\vec{k}} < \omega_{\rm F}.
		\label{eq: part2}
\end{align}
\end{enumerate}
Here, we define that $\hbar$ is the reduced Planck constant, $\beta=k_{\mathrm{B}}T$ is the inverse temperature, $T$ is the temperature, $k_{\mathrm{B}}$ is the Boltzmann constant, $\vec{k}$ is the wave number, $\omega_{\mathrm{D}}$ is the Debye frequency, and $\hbar\omega_{\vec{k}}= \hbar c_{\rm s} |\vec{k}| $ is the energy of the mode specified by $\vec{k}$.
By taking into account the temperature dependence of the Frenkel frequency $\omega_{\mathrm{F}}$, which is related to that of viscosity through the relation
\begin{align}
	\omega_{\mathrm{F}}=\omega_{\mathrm{M}} = \frac{G_{\infty}}{\eta},
\end{align}
we obtain from the partition function the total energy as
\begin{align} 
\langle\hat{H} \rangle=&N k_{\rm B} T \left(1+\frac{\alpha T}{2} \right) \notag \\
& \times  \left(3 D(x_{\rm D})-\left(\frac{\omega_{\rm F}}{\omega_{\rm D}} \right)^3 D(x_{\rm F}) \right),
\label{eq: enel_first}
\end{align}
where $D(x)$ is the Debye function
\begin{align}
D(x)\equiv \frac{3}{x^3}\int_0^x \frac{y^3 dy}{e^y -1},
\end{align}
$N$ is the number of particles in the system and $\alpha$ is the thermal expansion coefficient. 
The derivation of Eq. (\ref{eq: enel_first}) is explained in Supplemental Material. 
This expression depends on the ratio $\omega_{\mathrm{F}}/\omega_{\mathrm{D}}$, which reduces to $\omega_{\mathrm{F}}/\omega_{\mathrm{D}} \approx 0$ at the melting temperature,  and also reduces to $\omega_{\mathrm{F}}/\omega_{\mathrm{D}} \approx 1$ at the critical temperature by taking into account the temperature dependence of the Frenkel frequency \cite{Trachenko_2008, Bolmatov_2011,Bolmatov_2012, Trachenko_2016}. 
The heat capacity evaluated from Eq. (\ref{eq: enel_first}) quantitatively agree with the experimental date as shown in Fig. (\ref{fig:plot}).
Although the total energy obtained from our discussion is the same as in the phonon theory of liquid \cite{Trachenko_2008,Bolmatov_2011,Bolmatov_2012,Trachenko_2016}, the key idea in our approach is that the heat capacity reflects the change in the number of quasiparticles in each regime, which reproduces the proper behavior that interpolates between the number of modes at the melting point and at that of the critical point.
The decrease in heat capacity 
from $3Nk_{\mathrm{B}}$ to $2Nk_{\mathrm{B}}$ when lowering the temperature from the melting point to the critical point can be associated with the reduction of the number of quasiparticles from $3N$ to $2N$ modes. 
By understanding the dynamical modes via quasiparticles, we can derive the total energy much easier than the phonon theory of liquid \cite{Trachenko_2008,Bolmatov_2011, Bolmatov_2012, Trachenko_2016}.
We can also appropriately incorporate the contribution from hydrodynamic diffusion neglected in their theory \cite{Trachenko_2008,Bolmatov_2011, Bolmatov_2012, Trachenko_2016}.
This treatment, however, is not based on the single Hamiltonian or Lagrangian, which a physical system is expected to fundamentally possess.
In the following section, we present a possible way to construct a unified single Lagrangian which reproduces the number of quasiparticles at the melting temperature and at the critical point, and to interpolate between them. 

\section{Lagrangian theory for liquids}
For the purpose mentioned in the last section, we demonstrate in this section that the hydrodynamic Euler equation is equivalent to U(1) gauge theory with a specific gauge-fixing condition. We define a hydrodynamic Lagrangian based on this equivalence.
This Lagrangian constitutes the basis of our theory for heat capacity of liquids.
In addition, we add to the hydrodynamic Lagrangian a neutral scalar field to incorporate the elastic contribution explained in the last section.  
We start by considering the dynamics of the electromagnetic fields through the U(1) gauge Lagrangian 
\begin{align}
\mathcal{L}_A=-\frac{1}{4}F_{\mu\nu}F^{\mu\nu},
\label{eq: U1la}
\end{align}
where $F_{\mu\nu}=\partial_{\mu}A_{\nu}-\partial_{\nu} A_{\mu}$. In the present paper, we define 
$x^{\mu}=(ct,\vec{x}), x_{\mu}=\eta_{\mu\nu}x^{\nu}, A^{\mu}=( A^0,\vec{A})$ and adopt a notation $\eta^{\mu\nu}=\eta_{\mu\nu}={\rm diag}(1,-1,-1,-1), A_{\mu}=\eta_{\mu\nu}A^{\nu}$, where Greek indices run over space-time coordinates and the repeated greek index is summed over the space-time coordinates $0,1,2,3$. 
The Euler-Lagrange equation for this Lagrangian leads to the Maxwell's equations in vacuum 
\begin{align}
	\nabla \cdot \vec{E} = 0,\hspace{2ex}  \nabla \times \vec{B} -\frac{1}{c}\frac{\partial}{\partial t}\vec{E}  = 0,
\label{eq:Maxwell_equations}
\end{align}
where
\begin{align}
 \vec{E}=-\frac{1}{c}\frac{\partial}{\partial t} \vec{A}-\nabla A^0, \hspace{2ex}  
 \vec{B}= \nabla \times \vec{A},
 \label{eq:definitions_fields}
\end{align}
and $c$ is the speed of light. 
By combining Eqs. (\ref{eq:Maxwell_equations}) and  (\ref{eq:definitions_fields}), we obtain 
\begin{align}
&\frac{\partial }{\partial t}h +c^2 \nabla\cdot \vec{v}_{\parallel}=0,\
\label{eq: ma7}
\\
&\frac{\partial}{\partial t}\vec{v}_{\parallel} +\nabla h +ac\left( \frac{1}{c^2}\frac{\partial^2}{\partial t^2} -\nabla^2\right) \vec{A}=0,\
\label{eq: ma8}
\end{align}
where
\begin{align}
	h\equiv ac \nabla \cdot \vec{A},\hspace{2ex} \vec{v}_{\parallel} \equiv a \nabla A^0.
	\label{eq:definitions_fields_2}
\end{align}
where $a$ is the constant with appropriate dimension.
Now, we introduce a following gauge-fixing condition: 
\begin{align}
\nabla\cdot \left(\frac{1}{c^2}\frac{\partial^2}{\partial t^2} -\nabla^2\right) \vec{A} =0.\
\label{eq: fix1}
\end{align}
This gauge-fixing condition corresponds to the conservation of the vorticity in the hydrodynamics.
We give a detailed explanation for this gauge-fixing condition in Supplemental Material.
We also define $\vec{v}_{\perp}$ as 
\begin{align} 
\frac{\partial}{\partial t} \vec{v}_{\perp} \equiv 
ac \left(\frac{1}{c^2}\frac{\partial^2}{\partial t^2} -\nabla^2\right)\vec{A},\
\label{eq: fix2}
\end{align}
and we finally obtain the following equations: 
\begin{align}
	\frac{\partial}{\partial t} h + c^2 \nabla \cdot \vec{v} =0,
	\\
	\frac{\partial}{\partial t} \vec{v} + \nabla h = \vec{0},  
\end{align}
where $\vec{v} \equiv \vec{v}_{\parallel} + \vec{v}_{\perp} $. 
These equations are obviously equivalent to Eqs. (\ref{eq:equation_of_h}) and (\ref{eq:equation_of_v}) by replacing $c$ with $c_{\rm s}$. 
Therefore, we have establish the equivalence between hydrodynamics and  U(1) gauge theory with a specific gauge-fixing condition. 
From this equivalence, we can immediately define the hydrodynamic Lagrangian by Eq. (\ref{eq: U1la}) with Eq. (\ref{eq:definitions_fields_2}).
This hydrodynamic Lagrangian, however, cannot be applied to evaluate the thermodynamical property of liquids because the elastic behavior of liquids in high frequency regime, which was explained in the last section, is not incorporated yet.
To recover the contribution from the phonons in high frequency regime, we add the neutral scalar field $\phi$ with its mass $m_{\mathrm{F}}$, and we propose a Lagrangian for liquids as follows: 
\begin{align}
\mathcal{L}=\mathcal{L}_A+\frac{1}{2}\partial_{\mu}\phi \partial^{\mu}\phi -\frac{m_{\rm F}^2 c_{\rm s}^2}{2\hbar^2}\phi^2,\
\label{eq: tolag}
\end{align}
where $m_{\rm F} \equiv \hbar\omega_{\rm F} /c_{\rm s}^2$. 
The equations of motion derived from this Lagrangian are expressed as 
\begin{align}
&\partial_{\mu}F^{\mu\nu}=0,~~~\left(\partial_{\mu}\partial^{\mu} +m_{\rm F}^2 c_{\rm s}^2/\hbar^2 \right)\phi=0.
\label{eq:eq_original_1}
\end{align}
By adopting the Coulomb gauge, $A^0=0$ and $\partial_i A^i=0$, the equations of motion are rewritten as
\begin{align}
&\left(\frac{1}{c_{\rm s}^2} \frac{\partial^2}{\partial t^2} -\nabla^2 \right)A^i=0 ,\\
&\left(\frac{1}{c_{\rm s}^2} \frac{\partial^2}{\partial t^2} -\nabla^2 +\frac{m_{\rm F}^2 c_{\rm s}^2}{\hbar^2} \right)\phi=0.
\label{eq: eqmot}
\end{align}
In the high frequency regime, that is, the high energy regime $m_{\rm F} c_{\rm s}^2\ll \hbar \omega$, we can neglect the mass term of the scalar field $\phi$.
Here, we define
\begin{align}
(u_{\rm L})_i \propto \partial_i\phi,~~~(u_{\rm T})_i\propto A^i.
\label{eq:relation_elastic}
\end{align}
It is noted that the constant with appropriate dimension in Eq.(\ref{eq:relation_elastic}) can be recovered without difficulty.
We finally obtain the following equations:
\begin{align}
\left(\frac{1}{c_{\rm s}^2} \frac{\partial^2}{\partial t^2} -\nabla^2 \right)\vec{u}_{\rm L}=0,~~\left(\frac{1}{c_{\rm s}^2}\frac{\partial^2}{\partial t^2} -\nabla^2 \right)\vec{u}_{\rm T}=0,
\end{align}
where $\nabla\cdot \vec{u}_{\rm T}=0$ and $\nabla\times \vec{u}_{\rm L}=0$ are satisfied.
Consequently, Eq. (\ref{eq:eq_original_1}) becomes the wave equations for the longitudinal mode $\vec{u}_{\rm L}$ and  the transverse modes $\vec{u}_{\rm T}$.
On the other hand, in the low energy regime  $\hbar \omega \ll m_{\rm F} c_{\rm s}^2$, the scalar field $\phi$ becomes dynamically ineffective, because Eq. (\ref{eq: eqmot}) can be approximated as $(m_{\rm F}^2 c_{\rm s}^2/\hbar^2) \cdot \phi\simeq 0$, which results in the recovery of hydrodynamics in this  energy regime by adopting the gauge-fixing condition, Eq. (\ref{eq: fix1}).
It is noted that the Lagrangian, Eq. (\ref{eq: tolag}), can be transformed from the elastic regime to the hydrodynamic regime by choosing the gauge-fixing condition, which is the consequence of a gauge symmetry in this theory.  
Therefore, this Lagrangian appropriately reproduces the change in the number of modes depending on temperature between the melting point and the critical point.
The relative weight of each mode in the Lagrangian can be now controlled by the mass $m_{\rm F}$, which decides whether the field $\phi$ is dynamically effective or not depending on the temperature.    
Now, we discuss heat capacity of liquids based on the Lagrangian, Eq. (\ref{eq: tolag}) . This Lagrangian allows us to evaluate the partition function by means of a similar procedure to the Debye model or finite-temperature field theory in statistical physics \cite{Landau_1969_statistical1,Kapusta_Gale_2006}. 
As a result, we obtain a unified partition function
\begin{align}
Z_{\rm total}=\left(\prod_{\vec{k}}e^{-\beta E^A_{\vec{k}}}\right)^2 \left(\prod_{\vec{k}}e^{-\beta E_{\vec{k}}^{\phi}}\right),
\label{eq:partition_function_from_Lag}
\end{align}
where we define two energies of the quasiparticles, $E_{\vec{k}}^A=c_{\rm s} \hbar |\vec{k}|$ and $E_{\vec{k}}^{\phi}=\sqrt{(c_{\rm s} \hbar \vec{k})^2 +(m_{\rm F} c_{\rm s}^2)^2}$.
According to this partition function, we obtain the internal energy of liquids in equilibrium, which approximately reproduces Eq. (\ref{eq: enel_first}) up to the first order in
 $\beta \hbar\omega_{\rm F}$.
We give a detailed derivation of the heat capacity from the partition function, Eq. (\ref{eq:partition_function_from_Lag}),  in Supplemental Material.
As mentioned above, the expression obtained by our Lagrangian qualitatively reproduces the continuous decrease in the heat capacity from $3Nk_{\mathrm{B}}$ to $2Nk_{\mathrm{B}}$, and each limit corresponds to the number of quasiparticles at the melting point and at the critical point, respectively.
Our theoretical standpoint is the hydrodynamic equation as opposed to the phonon theory of liquid \cite{Trachenko_2008,Bolmatov_2011, Bolmatov_2012, Trachenko_2016}, which approaches the liquid phase from the solid phase. 
Our theory, however, does not need any estimation for the contributions to the total energy from each mode \cite{Trachenko_2008,Bolmatov_2011, Bolmatov_2012, Trachenko_2016}. 
A clear derivation of the heat capacity crucially originates from the quasiparticle description of liquids that specifies the number of modes at the melting temperature and at the critical temperature.   
A unified treatment of the hydrodynamic and the elastic behaviors with a correct interpolation between them is realized through the gauge symmetry which allows the natural transformation of the dynamical equation into each regime. 
We constructed a general, thermodynamically accessible model for liquids, which incorporates the qualitative behaviors at the melting point and at the critical point with natural interpolation.
This is as if the Debye model properly predicts the $T^3$-law at the lower temperature limit, and the Dulong-Petit law at the higher temperature limit with interpolation between two limiting cases.

\section{Conclusions}
In the present paper, we propose a thermodynamical model for evaluating the heat capacity of liquids.
We suggest a microscopic interpretation of the hydrodynamic equations and introduce a quasiparticle picture for the dynamics of particles in liquids. 
With the aid of a newly proposed equivalence between hydrodynamics and U(1) gauge theory, this interpretation allows us to construct a hydrodynamic Lagrangian whose thermodynamic property is easily calculated by appropriate procedures in statistical mechanics. 
By taking the elastic behavior in liquids into account by means of a neutral scalar field, we also propose a new Lagrangian for liquids as a possible way to calculate the heat capacity of liquids. 
It should be emphasized that the gauge symmetry of our theory enables us to establish a unified treatment of the hydrodynamic and the elastic equations, which appropriately reproduces the number of modes at the melting point and at the critical point.   
Our theory can be regarded as the counterpart in liquids to the Debye model in solids. 
In the future study, we will apply our theory to other thermodynamical properties of liquids.

\section*{Acknowledgement}
We would like to thank Toshihiro Kawakatsu, Motoi Endo, Naoto Kan, Kiyoharu Kawana, Ryota Kojima, Katsumasa Nakayama, Hikaru Ohta, Sayuri Takatori, and Sumito Yokoo for helpful discussions.


\bibliography{reference.bib} 

\clearpage
\begin{widetext}
\section*{Supplemental Material for ``Heat capacity of liquids in light of hydrodynamics as U(1) gauge theory"}

This supplemental material provides (I) the equivalence between hydrodynamics and $U(1)$ gauge theory and (II) the evaluation of the total energy of liquids.\\
\\

(I){\it The equivalence between hydrodynamics and $U(1)$ gauge theory.---}
In this section, we first discuss the possibility of our gauge-fixing condition.
In general, gauge vectors can be expressed as
\begin{align}
\vec{A}=\nabla \chi_A +\nabla\times \vec{X}_A.
\end{align}
In the equivalence between hydrodynamics and $U(1)$ gauge theory, we adopt following gauge-fixing condition:
\begin{align}
\partial_{\mu}\partial^{\mu}\chi_A=0.
\end{align}
Even if $\partial_{\mu}\partial^{\mu}\chi_A\neq 0$ is satisfied, we can obtain a suitable gauge vector ${A'}^{\mu}$ as
\begin{align}
\vec{A'}=\nabla \chi'_A +\nabla\times \vec{X}_A,~~~\partial_{\mu}\partial^{\mu}\chi'_A=0,
\end{align}
where we define as
\begin{align}
\chi'_A=\chi_A-\alpha,
\end{align}
by the gauge transformation:
\begin{align}
A^{\mu}\to {A'}^{\mu}=A^{\mu}+\partial^{\mu}\alpha.
\end{align}
Therefore, for a given gauge vector $A^{\mu}$, we can always choose as
\begin{align}
\alpha(x)=\chi_A(x)+\int d^3k \tilde{\alpha}(\vec{k})e^{ik\cdot x},
\end{align}
where $\tilde{\alpha}$ is an arbitrary function of $\vec{k}$ and we define $k^{\mu}=(c|\vec{k}|,\vec{k})$.
This means that our gauge-fixing is possible at all times.
Then, we obtain following relations:
\begin{align}
&\partial_{\mu}\partial^{\mu}\vec{A'}=\nabla\times\left(\partial_{\mu}\partial^{\mu}\vec{X}_A \right),
\\
&\nabla\cdot \vec{A'}=\nabla\cdot \nabla\chi'_A=\int d^3 k ~\vec{k}^2 \tilde{\alpha}(\vec{k})e^{ik\cdot x},
\end{align}
which is generally non-zero.
Thus, we define following quantities:
\begin{align}
&h\equiv ac \nabla\cdot \vec{A'},
\\
&\vec{v}_{\parallel}\equiv a \nabla {A'}^{0},
\\
&\frac{\partial}{\partial t}\vec{v}_{\perp}\equiv ac\partial_{\mu}\partial^{\mu}\vec{A'},
\end{align}
where $\nabla\times \vec{v}_{\parallel}=0$ and $\nabla\cdot \vec{v}_{\perp}=0$ are satisfied.
As the result, we can obtain the equations of the motion of the hydrodynamics from the equation of motion of the U(1) gauge theory.

(II){\it The evaluation of the total energy of liquids.---}
Let us evaluate the total energy of liquids.
We consider the total energy of liquids for partition functions, Eq.(\ref{eq: part1}) and Eq.(\ref{eq: part2}).
From these partition function, the energy of liquids in higher frequency regime is expressed as follows:
We consider the total energy of liquids for partition functions, Eq.(\ref{eq: part1}) and Eq.(\ref{eq: part2}).
\begin{align}
\langle \hat{H} \rangle\left(\omega_{\rm F} <\omega <\omega_{\rm D}\right)&=E_0(\omega_{\rm F} <\omega<\omega_{\rm D})+ \frac{3 V}{(2\pi)^3}\int_{\omega_{\rm F}}^{\omega_{\rm D}} d\omega 4\pi \frac{\omega^2}{c_{\rm s}^3}  \frac{\hbar \omega}{e^{\beta \hbar \omega}-1},
\label{eq: tenehi}
\end{align}
where $V$ is a volume of the system and $E_0(\omega_{\rm  F}<\omega<\omega_{\rm D})$ is the zero point energy:
\begin{align}
E_0(\omega_{\rm F}<\omega<\omega_{\rm D})= \frac{3 V}{(2\pi)^3}\int_{\omega_{\rm F}}^{\omega_{\rm D}} 4\pi \frac{\omega^2}{c_{\rm s}^3} d\omega \frac{\hbar \omega}{2}.
\end{align}
By using the Debye's function, we express Eq.(\ref{eq: tenehi}) as following form:
\begin{align}
\langle \hat{H} \rangle\left(\omega_{\rm F} <\omega <\omega_{\rm D}\right)=E_0(\omega_{\rm F}<\omega<\omega_{\rm D})+ \frac{V}{2 \pi^2 c_{\rm s}^3} k_{\rm B} T \left(\omega_{\rm D}^3 D(x_{\rm D}) -\omega_{\rm F}^3 D(x_{\rm F}) \right),
\label{eq:energy_high}
\end{align}
The Debye's function $D(x)$ is often expressed as 
\begin{align}
D(x)\equiv \frac{3}{x^3}\int_0^x \frac{y^3 dy}{e^y -1},
\end{align}
where we define as $x\equiv \beta \hbar \omega$.
Similar to the high frequency regime, we can evaluate the energy of liquid in the lower frequency regime as 
\begin{align}
\langle \hat{H} \rangle\left(0 <\omega <\omega_{\rm F}\right)&=E_0(0<\omega<\omega_{\rm F})+ \frac{2 V}{(2\pi)^3}\int_{0}^{\omega_{\rm F}} d\omega 4\pi \frac{\omega^2}{c_{\rm s}^3}  \frac{\hbar \omega}{e^{\beta \hbar \omega}-1},
\\
&=E_0(0<\omega<\omega_{\rm F})+ \frac{V}{3 \pi^2 c_{\rm s}^3} k_{\rm B} T D(x_{\rm F}),
\end{align}
where $E_0(0<\omega<\omega_{\rm F})$ is the zero point energy:
\begin{align}
E_0(0<\omega<\omega_{\rm F})=\frac{2 V}{(2\pi)^3}\int_{0}^{\omega_{\rm F}} 4\pi \frac{\omega^2}{c_{\rm s}^3} d\omega \frac{\hbar \omega}{2}.
\label{eq:energy_low}
\end{align}
Combining Eq.(\ref{eq:energy_high}) and Eq.(\ref{eq:energy_low}), we obtain the total energy of liquids:
\begin{align}
{\langle \hat{H} \rangle}&=\langle \hat{H} \rangle\left(0 <\omega <\omega_{\rm F}\right) +\langle \hat{H} \rangle\left(\omega_{\rm F} <\omega <\omega_{\rm D}\right),
\\
&=E_0 + V\frac{k_{\rm B}  T}{6\pi^2 c_{\rm s}^3} \omega_{\rm D}^3 \left(3 D(x_{\rm D}) -\left(\frac{\omega_{\rm F}}{\omega_{\rm D}}\right)^3 D(x_{\rm F}) \right),
\end{align}
where $E_0$ is the total zero-point energy, $E_0 (0<\omega<\omega_{\rm F})+E_0 (\omega_{\rm F}<\omega<\omega_{\rm D})$. 
If we take the heat expansion effect $V\to V(1+\alpha T/2)$ into account, we finally obtain
\begin{align} 
&\langle\hat{H} \rangle=\left(1+\frac{\alpha T}{2} \right) \left(E_0 +N k_{\rm B} T \left(3 D(x_{\rm D})-\left(\frac{\omega_{\rm F}}{\omega_{\rm D}} \right)^3 D(x_{\rm D}) \right) \right),\
\label{eq: enel_1}
\end{align}
where we used a relation, $N=V\omega_{\rm D}^3/6\pi^2 c_{\rm s}^3$ with the number of atoms $N$.
This result is consistent with \cite{Trachenko_2016}.

Next, we estimate the total energy of liquids from the partition function, Eq.(\ref{eq:partition_function_from_Lag}).
By definition of the total energy, we obtain as follows:
\begin{align}
\langle \hat{H}\rangle=\sum_{\vec{k}}\left[\left(\frac{E^{\phi}_{\vec{k}} }{2}+\frac{E^{\phi}_{\vec{k}}}{e^{\beta E^{\phi}_{\vec{k}} }-1} \right)  +2\cdot\left(\frac{E^A_{\vec{k}} }{2}+\frac{E^A_{\vec{k}}}{e^{\beta E^A_{\vec{k}} }-1} \right) \right].
\end{align}
%

%
For convenience, we define two energies, $\langle \hat{H}\rangle^{\phi}$ and $\langle \hat{H}\rangle^A$ as 
\begin{align}
&\langle \hat{H}\rangle^A=2\sum_{\vec{k}}\left[\frac{E^A_{\vec{k}} }{2}+\frac{E^A_{\vec{k}}}{e^{\beta E^A_{\vec{k}} }-1} \right],
\\
&\langle \hat{H}\rangle^{\phi}=\sum_{\vec{k}}\left[\frac{E^{\phi}_{\vec{k}} }{2}+\frac{E^{\phi}_{\vec{k}}}{e^{\beta E^{\phi}_{\vec{k}} }-1} \right].
\end{align}
Substituting summations to integrations, we obtain following forms:
\begin{align}
&\langle \hat{H}\rangle^A=E_0^A +\frac{2V}{(2\pi)^3}\int_0^{\omega_{\rm D}} d\omega~4\pi \frac{\omega^2}{c_{\rm s}^3}\frac{\hbar \omega}{e^{\beta \hbar \omega}-1},\
\label{eq: enA1}
\\
&\langle \hat{H}\rangle^{\phi}= E_0^{\phi} +\frac{V}{(2\pi)^3}\int_{\omega_{\rm F}}^{\omega_{\rm D}} d\omega ~ 4\pi \frac{\omega^2}{c_{\rm s}^3} \frac{\hbar \sqrt{\omega^2 -\omega_{\rm F}^2}}{e^{\beta \hbar \omega}-1}, \
\label{eq: enph1}
\end{align}
where $E_0^A$ and $E_0^{\phi}$ are the zero point energies:
Similar to the Debye's model in solids, in order to obtain a suitable interpolating function of the heat capacity of liquids which behaves as $3Nk_{\rm B} T$ and $2N k_{\rm B} T$ in the low and high temperature regime, respectively, we adopt the cut-off scale of the energy, $\omega_{\rm D}$ in Eq.(\ref{eq: enA1}) and Eq.(\ref{eq: enph1}). 
\begin{align}
&E_0^A=\frac{2 V}{(2\pi)^3}\int_0^{\omega_{\rm D}} d\omega 4\pi \frac{\omega^2}{c_{\rm s}^3}\frac{\hbar \omega}{2},
\\
&
E_0^{\phi}=\frac{ V}{(2\pi)^3}\int_{\omega_{\rm F}}^{\omega_{\rm D}} d\omega 4\pi \frac{\omega^2}{c_{\rm s}^3}\frac{\hbar \sqrt{\omega^2 -\omega_{\rm F}^2}}{2}.
\end{align}
By using the Debye's function, we express Eq.(\ref{eq: enA1}) as following form:
\begin{align}
&\langle \hat{H}\rangle^A=E_0^A+2V \frac{k_{\rm B} T}{6\pi^2 c_{\rm s}^3} \omega_{\rm D}^3 D(x_{\rm D}).\
\label{eq: enA2}
\end{align}
Besides, Eq.(\ref{eq: enph1}) is calculated as 
\begin{align} 
 \langle \hat{H}\rangle^{\phi}&=E_0^{\phi} +{\frac{V}{(2\pi)^3} \int_{\omega_{\rm F}}^{\omega_{\rm D}} d\omega 4\pi \frac{\omega^2}{c_{\rm s}^3} \frac{\hbar \omega \sqrt{1 -(\omega_{\rm F}/\omega)^2}}{e^{\beta \hbar \omega}-1}},
 \\
 &=E_0^{\phi} +\frac{V}{(2\pi)^3} \int_{\omega_{\rm F}}^{\omega_{\rm D}} d\omega 4\pi \frac{\omega^2}{c_{\rm s}^3} \frac{\hbar \omega}{e^{\beta \hbar \omega}-1}\left[1+\mathcal{O}((\hbar \omega_{\rm F}\beta)^2) \right],
 \\
 &=E_0^{\phi} +V \frac{k_{\rm B} T}{6\pi^2 c_{\rm s}^3} \omega_{\rm D}^3 \left(D(x_{\rm D}) -\left(\frac{\omega_{\rm F}}{\omega_{\rm D}}\right)^3 D(x_{\rm F})\right)+\mathcal{O}((\hbar \omega_{\rm F}\beta)^2).\
\label{eq: enph2}
\end{align} 
%
%
Combining Eq.(\ref{eq: enA2}) and Eq.(\ref{eq: enph2}), we obtain
%
%
\begin{align}
&\langle \hat{H} \rangle=E_0+V\frac{ k_{\rm B} T}{6\pi^2 c_{\rm s}^3}\omega_{\rm D}^3\left(3 D(x_{\rm D})-\left(\frac{\omega_{\rm F}}{\omega_{\rm D}} \right)^3 D(x_{\rm F}) +\mathcal{O}\left(( \hbar\omega_{\rm F}\beta)^2\right)\right)
,\
\label{eq: he1}
\end{align}
By taking the heat expansion effect into account, we obtain the total energy:
\begin{align} 
&\langle\hat{H} \rangle=\left(1+\frac{\alpha T}{2} \right) \left(E_0 +N k_{\rm B} T \left(3 D(x_{\rm D})-\left(\frac{\omega_{\rm F}}{\omega_{\rm D}} \right)^3 D(x_{\rm F})+\mathcal{O}\left(( \hbar\omega_{\rm F}\beta)^2\right) \right) \right),\
\label{eq: enel}
\end{align}
where $E_0$ is the zero-point energy, $E_0^A +E_0^{\phi}$.

\end{widetext}

\end{document}